\documentclass{article}
\usepackage{spconf,amsmath,graphicx}
\usepackage[linkcolor=black,urlcolor=blue,citecolor=black,pdfpagemode=None,pdfstartview=FitH,pdfview=FitH,colorlinks=true,pdftitle={NoLACE: Improving Low-Complexity Speech Codec Enhancement Through Adaptive Temporal Shaping},pdfauthor={Jan Buethe, Ahmed Mustafa, Jean-Marc Valin, Karim Helwani, Michael M. Goodwin}]{hyperref}

\usepackage{tikz}
\usetikzlibrary{shapes.geometric, arrows, backgrounds, calc, fit}

\definecolor{squidink}{RGB}{35, 47, 62}
\definecolor{anchor}{RGB}{0, 49, 129}
\definecolor{sky}{RGB}{32, 116, 213}
\definecolor{rind}{RGB}{251, 216, 191}
\definecolor{smile}{RGB}{255, 153, 0}

\tikzstyle{inout} = [rectangle, rounded corners, minimum width=2cm, minimum height=0.7cm, text centered, draw=black, fill=black!10]
\tikzstyle{conv1d} = [rectangle, rounded corners, minimum width=2.5cm, minimum height=0.7cm, text centered, draw=black, fill=smile!30]
\tikzstyle{pool} = [rectangle, rounded corners, minimum width=2.5cm, minimum height=0.7cm, text centered, draw=black, fill=sky!20]
\tikzstyle{tconv1d} = [rectangle, rounded corners, minimum width=2.5cm, minimum height=0.7cm, text centered, draw=black, fill=smile!70]
\tikzstyle{gru} = [rectangle, rounded corners, minimum width=2.5cm, minimum height=0.7cm, text centered, draw=black, fill=anchor!30]
\tikzstyle{adaconv1d} = [rectangle, rounded corners, minimum width=2.5cm, minimum height=0.7cm, text centered, draw=black, fill=smile!30]
\tikzstyle{adacomb1d} = [rectangle, rounded corners, minimum width=2.5cm, minimum height=0.7cm, text centered, draw=black, fill=smile!50]
\tikzstyle{adashape} = [rectangle, rounded corners, minimum width=1.5cm, minimum height=0.7cm, text centered, draw=black, fill=sky!60]
\tikzstyle{mathop} = [rectangle, rounded corners, minimum width=3cm, minimum height=0.7cm, text centered, draw=black, fill=rind!60]
\tikzstyle{op} = [circle, text centered, draw=black, fill=rind!60]
\tikzstyle{arrow}=[draw, -latex]

\newcommand\norm[1]{\lVert#1\rVert}

\title{NoLACE: Improving Low-Complexity Speech Codec Enhancement Through Adaptive Temporal Shaping}
\name{Jan B\"uthe, Ahmed Mustafa, Jean-Marc Valin, Karim Helwani, Michael M. Goodwin}
\address{Amazon Web Services\\ Palo Alto, USA \\
{\small \{jbuethe, ahdmust, jmvalin, helwk, mmg\}@amazon.com}}
\begin{document}
\ninept
\maketitle
\begin{abstract}
Speech codec enhancement methods are designed to remove distortions added by speech codecs. While classical methods are very low in complexity and add zero delay, their effectiveness is rather limited. Compared to that, DNN-based methods deliver higher quality but they are typically high in complexity and/or require delay. The recently proposed Linear Adaptive Coding Enhancer (LACE) addresses this problem by combining DNNs with classical long-term/short-term postfiltering resulting in a causal low-complexity model. A short-coming of the LACE model is, however, that quality quickly saturates when the model size is scaled up. To mitigate this problem, we propose a novel adatpive temporal shaping module that adds high temporal resolution to the LACE model resulting in the Non-Linear Adaptive Coding Enhancer (\mbox{NoLACE}). We adapt \mbox{NoLACE} to enhance the Opus codec and show that \mbox{NoLACE} significantly outperforms both the Opus baseline and an enlarged LACE model at 6, 9 and 12~kb/s. We also show that LACE and \mbox{NoLACE} are well-behaved when used with an ASR system.

\end{abstract}
\section{Introduction}
\label{sec:intro}

Degradation of speech through coding is a common problem in real-time communication scenarios where bandwidth is often limited and the speech codec therefore needs to operate below the transparency threshold. Improving such degraded speech has been a long-standing problem and a large variety of both classical and DNN based solutions have been proposed to address this issue \cite{chen_adaptive_postfiltering, zhao_enhancement, opus_resynthesis, gupta_speech_enhancement, korse_speech_enhancement, korse_postgan}.

While classical methods are very light-weight, DNN based methods tend to be either very complex or they require additional delay and often both is the case. This makes it difficult to deploy these methods for real-time applications on low-end devices like smart phones and it also makes it challenging to integrate them into switching codecs like Opus or EVS in which the speech codec is just one of many embedded modes.

The Linear Adaptive Coding Enhancer (LACE) \cite{lace} addresses these issues by combining the classical approach with a DNN which is used to calculate filter coefficients for long-term and short-term filters on a 5-ms-frame basis. These are then applied to the degraded signal in the classical way to enhance its harmonic structure and spectral envelope. The result is a very lightweight model that requires only 100~MFLOPS and is fully linear in the signal path. The model is furthermore fully causal on 20-ms frames and it is trained to be phase preserving. This allows for direct integration into existing communication codecs, where phase preservation ensures that seamless mode switching, as e.g. implemented in Opus or EVS, is maintained.

LACE has been shown \cite{lace} to significantly improve the Opus codec at 6, 9 and 12~kb/s and it provided about 60 \% of the MOS improvement of the non-causal LPCNet resynthesis method \cite{opus_resynthesis}, which requires 25~ms lookahead and comes with a complexity 3~GFLOPS.

A major drawback of the LACE model is, however, that quality quickly saturates when scaling up the model size, which essentially restricts it to be a very-low complexity tool with medium quality gain.

In this paper, we identify the low temporal resolution of LACE as the main cause for quality saturation. To mitigate this, we design an adaptive temporal shaping module, which calculates sample-wise gains on a frame basis using as input a feature vector and a temporal envelope of the signal to be shaped. We add the shaping module as a third custom DSP module to the LACE model and since the temporal shaping module also adds non-linear processing to the signal path we refer to the resulting model as Non-Linear Adaptive Coding Enhancer (\mbox{NoLACE}).

We adapt \mbox{NoLACE} as a multi-bitrate enhancer for the Opus speech coding mode\footnote{Demo samples are available at \href{https://282fd5fa7.github.io/NoLACE}{https://282fd5fa7.github.io/NoLACE}} and verify in a P.808 listening test that \mbox{NoLACE} significantly outperforms a LACE model of equivalent size at 6, 9 and 12~kb/s. The results show furthermore, that \mbox{NoLACE} at 12~kb/s scores close to the clean signal and that \mbox{NoLACE} at 6~kb/s achieves 92\% of the MOS improvement of the non-causal LPCNet resynthesis method.

As a second evaluation metric we test ASR performance by measuring the word error rates (WER) for Opus in combination with LACE, \mbox{NoLACE} and the LPCNet resynthesis method using the large SpeechBrain conformer model. The results show that LACE and \mbox{NoLACE} significantly improve WER at the lowest bitrate while the LPCNet resynthesis method leads to further degradation compared to the Opus baseline. At higher bitrates both LACE and \mbox{NoLACE} deliver WER close to the Opus baseline, which quickly converge to the WER of the clean signal.

Although we adapt \mbox{NoLACE} to enhance the Opus speech coding mode it can generally be used with any speech codec that provides explicit pitch information. Furthermore, the custom differentiable DSP (DDSP) modules used to build LACE and \mbox{NoLACE} can likely be used for other signal processing tasks. An implementation of the \mbox{NoLACE} model including a general implementation of the DDSP modules is available in the opus repository.\footnote{ \href{https://gitlab.xiph.org/xiph/opus/-/tree/icassp2024}{https://gitlab.xiph.org/xiph/opus/-/tree/icassp2024}}

The tested \mbox{NoLACE} model has a complexity of $\approx 620$ MFLOPS which requires only a small fraction of the compute capability available on common laptop or smart phone CPUs. Compared to fully neural codecs \cite{valin_lpcnet_coding, kleijn_lyra, zegidhour_soundstream,pia_nesc, jenrungrot_lmcodec, encodec, audiodec}, the approach of enhancing an existing codec also has the practical advantage of maintaining backward compatibility, leaving an inexpensive decoding option for low-end devices like microcontrollers.

\section{Proposed Model}
\subsection{Notation}
Throughout this paper we denote the clean signal by $x(t)$, the coded signal by $y(t)$ and the enhanced signal by $\hat y(t)$. Furthermore, we assume all these signals to be pre-emphasised with a factor 0.85. Furthermore, $n$ always denotes the sub-frame index of the Opus linear-predictive mode which operates on 5-ms subframes.

\subsection{Background: LACE}
The LACE model consists of two parts, a signal processing module and a feature encoder. The signal processing module implements two consecutive adaptive comb-filters, which make explicit use of a pitch lag $p_n$, followed by an adaptive convolution. The purpose of the feature encoder is to combine information from several input features into a latent feature vector $\varphi_n$, which is used in the signal processing module to compute filter coefficients on a 5-ms-frame basis. To avoid discontinuities, filter coefficients are interpolated on the first half of the 5-ms frames. Except for this input-independent interpolation, the system is essentially constant on individual frames, which restricts the temporal resolution 200~Hz.

For enhancing Opus, the input features are a mix of clean-signal features computed and quantized by the encoder (spectrum, pitch lag, ltp coefficients), noisy-signal features computed from the signal $y$ (cepstrum, auto-correlation) and bitrate information extracted at the decoder. For a detailed description of the input features we refer to \cite{lace}.

\subsection{NoLACE}
\begin{figure}
\center
\scalebox{0.75}{
\scalebox{1}{
\begin{tikzpicture}[node distance=1.2cm, scale=1]

\def\featurex{0}
\def\signalx{8}
\def\shapex{8.5}
\def\pitchx{6}

\node (features) [inout] at (0, 0) {Features};
\node (fconv1) [conv1d] at (0, -1.5) {Conv(k=1, s=1)};
\node (fpool) [pool] at (0, -3) {CPool(k=4, s=4)};
\node (fconv2) [conv1d] at (0, -4.5) {Conv(k=2, s=1)};
\node (ftconv) [tconv1d] at (0, -6) {TConv(k=4, s=4)};
\node (fgru) [gru] at (0, -7.5) {GRU};
\node (phi) at (2, -7.5) {\small$\varphi^{(1)}_n$};

\node at (0.75, -2.25) {$\#c=N_r$};
\node at (0.75, -3.75) {$\#c=N_h$};
\node at (0.75, -5.25) {$\#c=N_h$};
\node at (0.75, -6.75) {$\#c=N_h$};

\node (ftrans1) [conv1d] at (4, -3) {Conv(k=2)};
\node (phi2) at (6, -2.7) {\small$\varphi^{(2)}_n$};
\node (ftrans2) [conv1d] at (4, -4.5) {Conv(k=2)};
\node (phi3) at (6, -4.2) {\small$\varphi^{(3)}_n$};
\node (ftrans3) [conv1d] at (4, -6) {Conv(k=2)};
\node (phi4) at (6, -5.7) {\small$\varphi^{(4)}_n$};
\node (ftrans4) [conv1d] at (4, -9) {Conv(k=2)};
\node (phi5) at (6, -8.7) {\small$\varphi^{(5)}_n$};
\node (ftrans5) [conv1d] at (4, -12) {Conv(k=2)};
\node (phi6) at (6, -11.7) {\small$\varphi^{(6)}_n$};

\node (lags) at (\pitchx, 0) {$p_n$};

\node (sigin)  at (\signalx, 0) {$y(t)$};
\node (adacomb1) [adacomb1d] at (\signalx, -1.5) {AdaComb1};
\node (adacomb2) [adacomb1d] at (\signalx, -3) {AdaComb2};
\node (adaconv) [adaconv1d] at (\signalx, -4.5) {AdaConv1};
\node (adashape1) [adashape] at (\shapex, -6) {AdaShape1};
\node (adaconv2) [adaconv1d] at (\signalx, -7.5) {AdaConv2};
\node (adashape2) [adashape] at (\shapex, -9) {AdaShape2};
\node (adaconv3) [adaconv1d] at (\signalx, -10.5) {AdaConv3};
\node (adashape3) [adashape] at (\shapex, -12) {AdaShape3};
\node (adaconv4) [adaconv1d] at (\signalx, -13.5) {AdaConv4};
\node (denoised)  at (\signalx, -15) {$\hat y(t)$};

\draw [arrow, dashed] (features) -- (fconv1);
\draw [arrow, dashed] (fconv1) -- (fpool);
\draw [arrow, dashed] (fpool) -- (fconv2);
\draw [arrow, dashed] (fconv2) -- (ftconv);
\draw [arrow, dashed] (ftconv) -- (fgru);

\draw [arrow, dashed] (fgru) -- (phi);
\draw [arrow, dashed] (phi) -- (2, -1.5) -- (adacomb1);
\draw [arrow, dashed] (4, -1.5) -- (ftrans1);
\draw [arrow, dashed] (ftrans1) -- (adacomb2);
\draw [arrow, dashed] (6, -3) -- (6, -3.75) -- (4, -3.75) -- (ftrans2);
\draw [arrow, dashed] (ftrans2) -- (adaconv);
\draw [arrow, dashed] (6, -4.5) -- (6, -5.25) -- (4, -5.25) -- (ftrans3);
\draw [arrow, dashed] (ftrans3) -- (adashape1);
\draw [arrow, dashed] (6, -6) -- (6, -7.5) -- (4, -7.5) -- (ftrans4);
\draw [arrow, dashed] (6, -7.5) -- (adaconv2);
\draw [arrow, dashed] (ftrans4) -- (adashape2);
\draw [arrow, dashed] (6, -9) -- (6, -10.5) -- (4, -10.5) -- (ftrans5);
\draw [arrow, dashed] (6, -10.5) -- (adaconv3);
\draw [arrow, dashed] (ftrans5) -- (adashape3);
\draw [arrow, dashed] (6, -12) -- (6, -13.5) -- (adaconv4);

\draw [arrow] (sigin) -- (adacomb1);
\draw [arrow] (adacomb1) -- (adacomb2);
\draw [arrow] (adacomb2) -- (adaconv);

\draw [arrow] ($(adaconv.south) - (0.5, 0)$) -- ($(adaconv2.north) - (0.5, 0)$);
\draw [arrow] ($(adaconv2.south) - (0.5, 0)$) -- ($(adaconv3.north) - (0.5, 0)$);
\draw [arrow] ($(adaconv3.south) - (0.5, 0)$) -- ($(adaconv4.north) - (0.5, 0)$);

\draw [arrow] ($(adaconv.south) + (0.5, 0)$) -- ($(adashape1.north)$);
\draw [arrow] (adashape1.south) -- ($(adaconv2.north) + (0.5, 0)$);
\draw [arrow] ($(adaconv2.south) + (0.5, 0)$) -- ($(adashape2.north)$);
\draw [arrow] (adashape2.south) -- ($(adaconv3.north) + (0.5, 0)$);
\draw [arrow] ($(adaconv3.south) + (0.5, 0)$) -- ($(adashape3.north)$);
\draw [arrow] (adashape3.south) -- ($(adaconv4.north) + (0.5, 0)$);

\draw [arrow] (adaconv4) -- (denoised);

\draw [arrow, dashed] (features) -- (lags);
\draw [arrow, dashed] (lags) -- (\pitchx, -0.8) -- (7, -0.8) -- (7, -1.2);
\draw [arrow, dashed] (\pitchx, -0.8) -- (\pitchx, -2.2) -- (7, -2.2) -- (7, -2.7);

\begin{pgfonlayer}{background}
\filldraw [line width=4mm, join=round, black!10]
(fconv1.north -| fconv1.east) rectangle (fgru.south -| fgru.west)
(adacomb1.north -| ftrans1.west) rectangle (adaconv4.south -| adashape3.east);
\end{pgfonlayer}{background}

\end{tikzpicture}
}
}
\caption{High-level overview of the \mbox{NoLACE} model. The feature encoder which transforms the input features into a latent feature vector $\varphi_n^{(1)}$ is depicted on the left and the number of channels are indicated with $\#c=$. The signal processing unit on the right applies first a series of two comb-filtering and spectral shaping operation before entering a select-shape-mix iteration involving the proposed adaptive temporal shaping modules AdaShape}\label{f:NoLACE}
\end{figure}
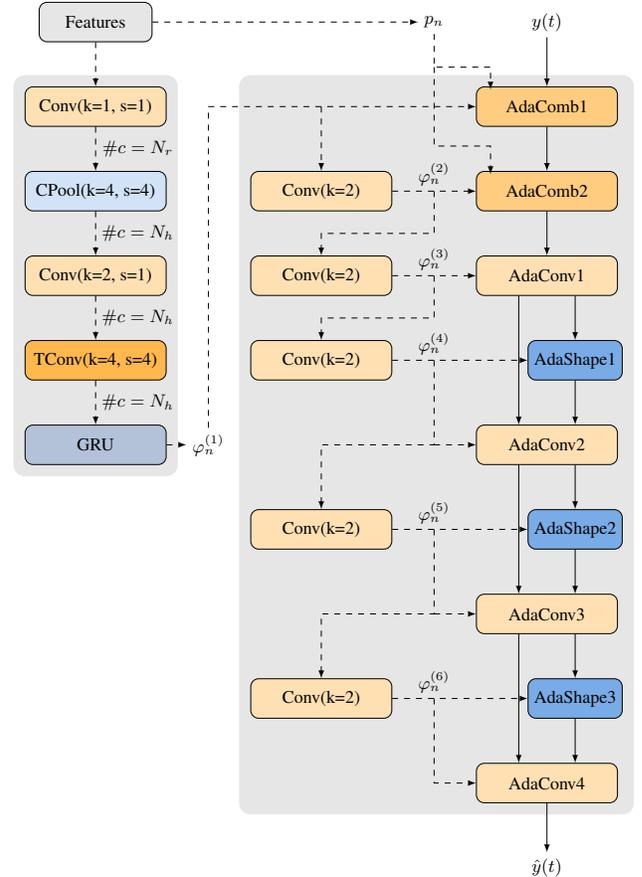

The \mbox{NoLACE} model has the same basic design as LACE. It consists of a feature encoder identical to the LACE feature encoder and a signal processing module displayed in Figure \ref{f:NoLACE}. The first part of the signal processing module follows the design of LACE, having two consecutive adaptive comb-filter modules (AdaComb) and one adaptive convolution module (AdaConv). What follows next is an iteration of a select-shape-mix procedure that is build around the adaptive temporal shaping module (AdaShape): the AdaConv1 module produces two output channels (select), one passing through the adaptive temporal shaping module AdaShape1 (shape) and one bypassing it. The two channels are then mixed together by the AdaConv2 module which, producing two output channels, instantaneously performs the selection operation for the next iteration. The reasoning behind this is the following: the initial AdaConv module implements a spectral shaping and as such selects the frequency components that are to be shaped by the AdaShape module. Since the AdaShape module is non-linear in the signal path, the selection also serves as aliasing control. The bypass channel carries complementary signal parts that are required to reconstruct the signal in the final mixing operation. The select-shape-mix procedure is carried out three times with AdaConv4 producing the final enhanced signal $\hat y(t)$.

A second addition to the signal processing module are additional feature transformations that are used to filter the latent feature vector while handing it down from layer to layer. While the main quality improvement for \mbox{NoLACE} comes from the temporal shaping modules, this additional filtering has been found to provide an additional small but significant quality improvement.

\mbox{NoLACE} and LACE share the same hyper parameters $N_r$ (the reduced feature dimension) and $N_h$ (the number of hidden channels, the GRU size and the dimension of the latent feature vectors $\varphi_n^{(k)}$). In fact, LACE can be recovered from \mbox{NoLACE} as the first output channel of the AdaConv1 module in Figure \ref{f:NoLACE} when setting $\varphi_n^{(3)} := \varphi_n^{(2)} := \varphi_n^{(1)}$. While the LACE was trained with $N_r=96$ and $N_h=128$ in \cite{lace}, we choose $N_r=96$ and $N_h=256$ for \mbox{NoLACE}. We similarly increase $N_h$ for the LACE comparison model in this paper.

\subsection{The Adaptive Temporal Shaping Module}
\begin{figure}
\center
\scalebox{0.75}{
\scalebox{1}{
\begin{tikzpicture}[node distance=1.2cm, scale=1]

\def\mwidth{3.1}

\node (signal) at (0, 2) {$\xi(t)$};
\node (absval) [mathop, minimum width=\mwidth cm] at (0, 0) {$\mid \cdot \mid$};
\node (avgpool) [pool, minimum width=\mwidth cm] at (0, -1.5) {AvgPool(k=4,s=4)};
\node (reshape1) [pool, minimum width=\mwidth cm] at (0, -3) {Frame($N/4$)};
\node (log) [mathop, minimum width=\mwidth cm] at (0, -4.5) {($\log(\cdot) - \mu_n, \mu_n)$};
\node (concat) [mathop, minimum width=\mwidth cm] at (0, -6) {Concat($\cdot$, $\varphi_n$)};
\node (conv1) [conv1d, minimum width=\mwidth cm] at (0, -7.5) {LReLU(Conv1d(k=2))};
\node (conv2) [conv1d, minimum width=\mwidth cm] at (0, -9) {exp(Conv1d(k=2))};
\node (reshape2) [pool, minimum width=\mwidth cm] at (0, -10.5) {Flatten};
\node (phi) at (-2, 2) {$\varphi_n$};
\node (mult) [op] at (3, -10.5) {$\ast$}; 
\node (output) at (3, -12) {$\xi_{\text{shaped}}(t)$};

\node (alpha) at (2.1, -10.2) {$\alpha(t)$};

\draw[arrow, dashed] (phi) -- (-2, -6) -- (concat);
\draw[arrow] (signal) -- (absval);
\draw[arrow] (absval) -- (avgpool);
\draw[arrow] (avgpool) -- (reshape1);
\draw[arrow] (reshape1) -- (log);
\draw[arrow] (log) -- (concat);
\draw[arrow] (concat) -- (conv1);
\draw[arrow] (conv1) -- (conv2);
\draw[arrow] (conv2) -- (reshape2);
\draw[arrow] (reshape2) -- (mult);
\draw[arrow] (0, 0.75) -- (3, 0.75) -- (mult);
\draw[arrow] (mult) -- (output);

\begin{pgfonlayer}{background}
\filldraw [line width=4mm, join=round, black!10, inner sep=1cm]
(-2.5, 1) rectangle (3.5, -11);
\end{pgfonlayer}{background}

\end{tikzpicture}
}
}
\caption{Adaptive temporal shaping module. Shapes are given in channels last format, $N$ denotes the frame size and $\mu$ denotes the frame-wise mean value.}\label{f:adashape}
\end{figure}
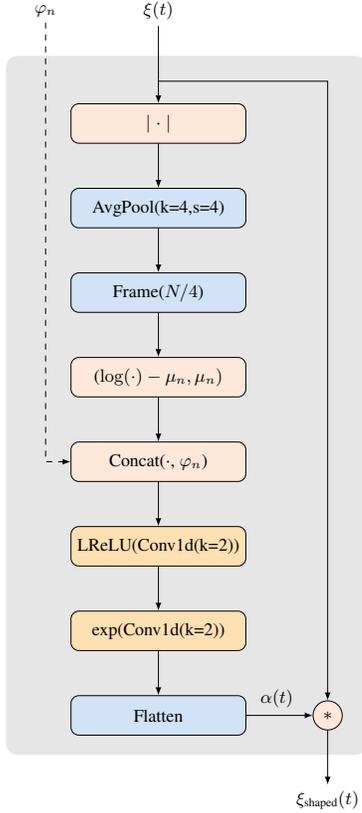

The adaptive temporal shaping module (AdaShape) modifies the input signal by multiplying each sample by an individual gain. These sample-wise gains are calculated on a frame basis from a temporal envelope, derived from the input signal as the average absolute values on 4-sample blocks, and a feature vector $\varphi_n$. To reduce the dynamic range, modelling is done in log-domain. To this end the temporal envelope is transformed to log domain and the frame-mean value $\mu_n$ is subtracted. The resulting zero-mean envelope is then concatenated with the mean value $\mu_n$ and the given feature vector $\varphi_n$ which are then passed through a two-layer convolutional network to derive the sample-wise gains in log domain as illustrated in Figure \ref{f:adashape}.

\subsection{Multi-Channel Adaptive Convolutions}
The select-shape-mix iteration requires adaptive convolutions with two channels. Since the AdaConv module was only defined for a single input and output channel in \cite{lace}, we extend the definition to an arbitrary number of input and output channels. This is done in analogy to regular 1d convolutions but kernel normalization and gain computation require some attention. In the original definition the single channel impulse response is calculated from a feature vector $\varphi_n$ as product of a shape
\begin{equation}
\kappa_n = \frac{W_\kappa \varphi_n + b_\kappa}{\norm{W_\kappa \varphi_n + b_\kappa}_2}
\end{equation}
and a gain
\begin{equation}\label{e:gain}
g_n = \exp(\alpha \tanh(W_g \varphi_n + b_g))
\end{equation}
to derive the impulse response for a single frame
\begin{equation}
h_n(\tau) = g_n \, \kappa_n(\tau),
\end{equation}
where $\tau$ denotes the filter tap index. The impulse responses are then interpolated on the first half of the frame to provide smooth transitions.

We extend this definition first to $m_1$ input channels and a single output channel by defining it as the sum of $m_1$ adaptive convolutions, where the $j$-th kernel shape is given by
\begin{equation}
    \kappa_n^{(j)} = \frac{W_\kappa^{(j)} \varphi_n + b_\kappa^{(j)}}{\sum_{i=1}^m \norm{W_\kappa^{(i)} \varphi_n + b_\kappa^{(i)}}_2}
\end{equation}
and where the kernel gain is given by \eqref{e:gain} for all $m_1$ adaptive convolutions, i.e. we jointly normalize over all input channels and use a shared gain. An adaptive convolution with $m_1$ input channels and $m_2$ output channels is defined by concatenating $m_2$ adaptive convolutions with $m_1$ input channels and a single output channel along the channel dimension.

\section{Training}

We trained on 165 hours of clean speech sampled at 16~kHz which are collected from multiple high-quality TTS datasets \cite{demirsahin-etal-2020-open, kjartansson-etal-2020-open, kjartansson-etal-tts-sltu2018, guevara-rukoz-etal-2020-crowdsourcing, he-etal-2020-open, oo-etal-2020-burmese, van-niekerk-etal-2017, gutkin-et-al-yoruba2020, bakhturina21_interspeech} containing more than 900 speakers in 34 languages and dialects. The input signals $y$ were generated using a patched version of libopus that restricts Opus to linear-predictive mode and wideband encoding. Furthermore, the encoder is modified to randomly switch encoding parameters bitrate, complexity and packet\_loss\_percent every 249-th frame.\footnote{\href{https://gitlab.xiph.org/xiph/opus/-/tree/exp-neural-silk-enhancement}{https://gitlab.xiph.org/xiph/opus/-/tree/exp-neural-silk-enhancement}}

\subsection{Model Pre-Training}
In a first step, \mbox{NoLACE} is pre-trained using the same combination of regression losses, $10 \, \mathcal{L}_{\mathrm{phase}}  + 2\, \mathcal{L}_{\mathrm{env}} + \mathcal{L}_{\mathrm{spec}},$ from \cite{lace}. Pre-training is carried out for 50 epochs using the Adam optimizer with $\beta_1=0.9$ and $\beta_2=0.999$, a sequence length of $0.5$ seconds, a batch size of 256 and a learning rate decay factor of $2.5\times 10^{-5}$.

\subsection{Adversarial Training}
For adversarial training we use spectrogram discriminators at multiple resolutions as proposed in \cite{univnet}. We follow the setup in \cite{fwgan} and use 6 STFT discriminators $D_k$ but apply a few modifications: First, each discriminator takes as input a log-magnitude spectrogram calculated from size-$2^{k+5}$ STFTs with 75\% overlap. By abuse of notation we still write $D_k(x)$ or $D_k(\hat y)$, treating the log-magnitude STFT transform as part of the discriminator. We furthermore apply strides along the frequency axis to keep the frequency range of the receptive fields constant. This has been found to increase the ability of discriminators with high frequency resolution to detect inter-harmonic noise. Finally, we concatenate a two-dimensional frequency positional sine-cosine embedding to the input channels of every 2d-convolutional layer.

We train \mbox{NoLACE} as a least-squares GAN \cite{lsgan}. First we note that the coded input signal $y$ depends only on the clean signal $x$ and the encoder parameters $\pi_\mathrm{enc}$.  With this we define the adversarial part of the training loss for \mbox{NoLACE} as
\begin{equation}
    \mathcal{L}_{\textrm{adv}}(x, \hat y) = \frac{1}{6}\sum_{k=1}^6 E_{x, \pi_\mathrm{enc}}[(1 - D_k(\hat y))^2] + \mathcal{L}_{\mathrm{feat}}(D_k, x, \hat y),
\end{equation}
where $\mathcal{L}_{\mathrm{feat}}$ denotes the standard feature matching loss, i.e. the mean of the $L^1$ losses of hidden layer outputs for $x$ and $\hat y$.

For regularization and to maintain phase preservation we also add the following combination of pre-training losses
\begin{equation}
    \mathcal{L}_{\mathrm{reg}} := \frac{60}{155} \mathcal{L}_{\mathrm{env}} + \frac{30}{155} \mathcal{L}_{\mathrm{phase}} + \frac{3}{155} \mathcal{L}_{\mathrm{spec}},
\end{equation}
whence the final training loss for \mbox{NoLACE} is given by
\begin{equation}
    \mathcal{L}_{\textrm{NoLACE}}(x, \hat y) = \mathcal{L}_{\textrm{adv}}(x, \hat y) + \mathcal{L}_{\mathrm{reg}}(x, \hat y).
\end{equation}

Simultaneously, the discriminators are trained to minimize the losses
\begin{equation}
    \mathcal{L}_{D_k}(x, \hat y) = E_{x, \pi_\mathrm{enc}}[D_k(\hat y)^2 + (1 - D_k(x))^2].
\end{equation}

Adversarial training is carried out for another 50 epochs with a fixed learning rate of $10^{-4}$ and a batch size of 64, which corresponds to roughly 930~K training steps. We used the Adam optimizer with $\beta_1=0.9$ and $\beta_2=0.999$ for both \mbox{NoLACE} and the discriminators.

\section{Evaluation}
\subsection{Listening Test}
We evaluated the quality of \mbox{NoLACE} using 192~clean English speech clips from the NTT Multi-Lingual Speech Database for Telephonometry, which was not included in the training data. Using the crowd-sourcing methodolgy from ITU-R P.808\cite{p808}, we tested Opus, Opus + LACE and Opus + \mbox{NoLACE} at 6, 9, and 12~kb/s.

For a fair comparison we trained both LACE and \mbox{NoLACE} with $N_r=96$ and $N_h=256$. This leads to a complexity of 280~MFLOPS with 900~K parameters for LACE (roughly a 3x increase) and a complexity of 620~MFLOPS with 1.8~M parameters for \mbox{NoLACE}. No adversarial training was performed for the LACE model since it was found to degrade quality.

We furthermore included the LPCNet resynthesis method from \cite{opus_resynthesis} at 6~kb/s as an additional test point. The method adds 25~ms delay to the decoder and is therefore not a realistic replacement option for LACE or \mbox{NoLACE} so it rather serves as an interesting reference point.

The results show that \mbox{NoLACE} consistently outperforms the large LACE model. Furhtermore, \mbox{NoLACE} achieves roughly 92\% of the MOS improvement of LPCNet resynthesis\cite{opus_resynthesis}. LACE, on the other hand, achieves about $66\%$ of MOS improvement, which is 6~pp higher than the results reported for the smaller LACE model in \cite{lace}. This indicates that quality saturates indeed quickly when enlarging the model size.

\begin{figure}
\begin{center}
\includegraphics[scale=0.8]{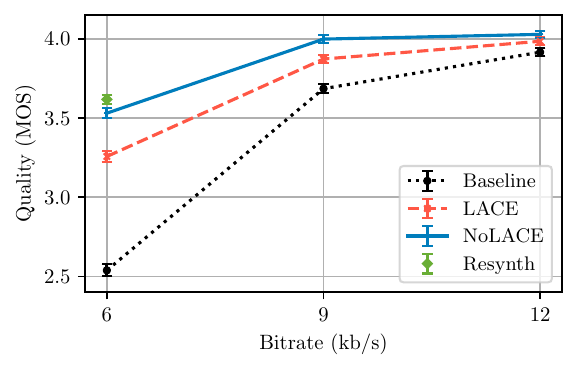}
\caption{P.808 results. The clean signal has a MOS of $4.06\pm 0.025$. LACE consistently outperforms the baseline and \mbox{NoLACE} consistently outperforms LACE at all bitrates. At 6~kb/s, \mbox{NoLACE} achieves $92 \%$ of the MOS improvement of the LPCNet resynthesis method which requires 25ms delay and 5x the complexity of \mbox{NoLACE}.}
\end{center}
\end{figure}

\subsection{ASR testing}
We evaluated the impact of Opus and the three enhancement methods LACE, \mbox{NoLACE} and LPCNet resynthesis on ASR performance of the large SpeechBrain \cite{speechbrain} conformer ASR model\footnote{\href{https://huggingface.co/speechbrain/asr-conformer-transformerlm-librispeech}{https://huggingface.co/speechbrain/asr-conformer-transformerlm-librispeech}} using the native clean speech test set of the LibriSpeech ASR corpus \cite{librispeech}. The results in Table \ref{t:asr-wer} show that Opus coding has a significant impact on ASR performance at very low bitrates, increasing WER by roughtly one pp at 6~kb/s. While the resynthesis method further increases WER by approx. 0.2~pp at this bitrate, both LACE and \mbox{NoLACE} enhancement leads to an improvement of approx. 0.6 resp 0.5~pp indicating that LACE and \mbox{NoLACE} make up for about half the errors introduced by Opus coding.

For higher bitrates, Opus, LACE and \mbox{NoLACE} quickly converge to the clean speech WER indicating that LACE and \mbox{NoLACE} can be used with an ASR system without having to retrain.

\begin{table}
\begin{center}
\begin{tabular}{l|cccc}
condition           & 6~kb/s & 9~kb/s & 12~kb/s & 20~kb/s\\
\hline
clean               & 2.01\% & 2.01\% & 2.01\% & 2.01\%\\
Opus                & 3.08\% & 2.15\% & 2.07\% & 2.03\%\\
Opus + LACE         & 2.46\% & 2.13\% & 2.05\% & 2.03\% \\
Opus + NoLACE       & 2.56\% & 2.18\% & 2.06\% & 2.02\%\\
Opus + resynthesis  & 3.26\% & - & - & -\\
\end{tabular}
\caption{Word error rates for Opus in combination with different enhancement methods. At the lowest bitrate WER is significantly increased for Opus condition. LACE and \mbox{NoLACE} significantly reduce WER for this bitrate while the LPCNet resynthesis method further increases it. At higher bitrates WER for LACE and \mbox{NoLACE} quickly reduce and are close to Opus WER.}\label{t:asr-wer}
\end{center}
\end{table}

\section{Conclusion}
In this paper we identified low temporal resolution as the main bottleneck of the LACE model for scaling to higher quality. We introduced the adaptive temporal shaping module (AdaShape), used it to design the Non-Linear Adaptive Coding Enhancer (NoLACE) and demonstrateed in a P.808 listening test that \mbox{NoLACE} consistently outperforms LACE. Furthermore, we conducted an ASR test which showed that ASR performance is maintained, and at low bitrates even improved, when adding LACE or \mbox{NoLACE}. The model is causal and phase-preserving and can be implemented with insignificant complexity overhead on common smart phone CPUs, which allows for direct integration into codecs with dedicated speech-coding mode. We believe such a low-complexity enhancement algorithm will be most useful for enhancing the quality of existing classical speech codecs while maintaining compatibility.

\section{Acknowledgment}
We would like to thank Minho Jin for assisting with the ASR test.

\bibliographystyle{IEEEbib}
\bibliography{audio}

\end{document}